\documentclass[aps,prb, twocolumn,showpacs,preprintnumbers,superscriptaddress]{revtex4}

\usepackage{graphicx}
\usepackage{dcolumn}
\usepackage{bm}

\begin{document}

\title{Suppression of the low-temperature phase-separated state under pressure  in (Eu$_{1-x}$Gd$_{x}$)$_{0.6}$Sr$_{0.4}$MnO$_{3}$ ($x=0,0.1$)}

\author{Tasuku Inomata}
\author{Michiaki Matsukawa} 
\email{matsukawa@iwate-u.ac.jp }
\author{Daichi Kimura} 
\author{Yoshiaki Yamato} 
\author{Satoru Kobayashi} 
\affiliation{Department of Materials Science and Engineering, Iwate University , Morioka 020-8551 , Japan }
\author{Ramanathan Suryanarayanan}
\affiliation{Laboratoire de Physico-Chimie de L'Etat Solide,CNRS,UMR8648
 Universite Paris-Sud, 91405 Orsay,France}
\author{Sigeki Nimori}
\affiliation{National Institute for Materials Science, Tsukuba 305-0047 ,Japan}
\author{Keiichi Koyama}
\affiliation{Graduate School of Science and Engineering,Kagoshima University, Kagoshima 890-0065, Japan}
\author{Kohki Takahashi}
\author{Kazuo Watanabe}
\author{Norio Kobayashi}
\affiliation{Institute for Materials Research, Tohoku University, Sendai  
980-8577, Japan}

\date{\today}

\begin{abstract}
We have demonstrated the effect of pressure on the steplike metamagnetic transition and its associated magnetostriction in (Eu$_{1-x}$Gd$_{x}$)$_{0.6}$Sr$_{0.4}$MnO$_{3}$ ($x=0$ and 0.1). The critical field initiating the field induced ferromagnetic transition in both samples is lowered by the applied pressure.
The further application of external pressure up to 1.2 GPa on the $x=0$ parent sample causes a spontaneous ferromagnetic transition with a second-oder like character, leading to collapses of the steplike transition and its concomitant lattice striction.
These findings indicate a crucial role of the low-temperature phase separated state characterized by a suppressed magnetization upon decreasing temperature.
 

\end{abstract}

\renewcommand{\figurename}{Fig.}
\maketitle
\section{INTRODUCTION}
Manganites (RE$_{1-x}$AE$_{x}$MnO$_{3}$ ; RE=Y,rare earth; AE=Pb,Ca,Sr,Ba) with  perovskite structure have been extensively investigated since the discovery of colossal magnetoresistance (CMR) effect.\cite{TO00} 
The spontaneous insulator to metal transition and its associated CMR effect are well explained on the basis of the double exchange (DE) model between Mn$^{3+}$ and Mn$^{4+}$ ions.\cite{ZE51} Furthermore, the phase separation (PS) model, where the ferromagnetic (FM) metallic and antiferromagnetic charge ordered (AFM-CO) insulating clusters of competing electronic phases coexist, strongly supports experimental studies of manganites.\cite{DA01}
In a recent study\cite{DEM08} on the bandwidth-temperature-magnetic field phase diagram of  RE$_{0.55}$Sr$_{0.45}$MnO$_{3}$ with nearly half doping, the bandwidth was controlled by chemical substitution and hydrostatic pressure. In particular, near a critical pressure, the character of the ferromagnetic transition varies from the first to second order.
The applied pressure gives rise to a larger bandwidth accompanied by a decrease of short range charge/orbital ordering fluctuations, resulting in weakened first order transition.

Several recent researches for metamagnetic transitions of doped manganites in the phase separated state have revealed that ultrasharp magnetic transition of Pr$_{0.5}$Ca$_{0.5}$Mn$_{0.95}$Co$_{0.05}$O$_{3}$ appears at low temperatures. 
To account for this phenomenon, a martensitic model based on lattice mismatch between competing CO and FM phases has been proposed\cite{MAH02,GHI04,FIS04} but questions have been raised as to the validity of such a scenario.\cite{MA07}
It is believed that the martensitic scenario is ruled out since grain boundaries in polycrystalline samples act as "fire wall" against such a progress of magnetic transition, resulting in a suppression of its steplike transition.
Ghivelder et al., have observed a sudden temperature rise ($\sim$ 30 K)   associated with the steplike magnetic transition of La$_{5/8-x}$Pr$_{x}$Ca$_{3/8}$MnO$_{3}$ ($x=0.4$),
indicating a release of  large latent heat  (or entropy)  from low-temperature phase-separated to FM ordered states.  
According to their findings,\cite{GHI05,SAC06} after zero-field cooling, the manganite system  showing the steplike transition reaches a thermally stable state with a small, and almost time independent fraction of FM, which is dispersed as isolated regions within a CO matrix. This phase separated state is realized at low temperatures  by large energy barriers separating the AFM-CO and FM phases and strains due to the lattice mismatch between the coexisting phases. Increasing the temperature under the applied field, the system is then transfered from the equilibrium to non equilibrium states, promoting a growth of the FM phase over the majority AFM-CO matrix.

A close relationship of the $TH$ phase diagram with steplike metamagnetic transition for low band width manganites, Sm$_{0.55}$Sr$_{0.45}$MnO$_{3}$ and Eu$_{0.58}$Sr$_{0.42}$MnO$_{3}$, has been pointed out in a previous work.\cite{KA09} The negative curvature in the temperature dependence of critical magnetic field , $dH_{c}/dT<0$, at low temperatures is an essential condition for the magnetic transition observed  in doped manganites, where $ H_{c}$ is a critical field causing the ferromagnetic transition.
The FM transition of local region  generates  a release of latent heat, which drives its neighbor regions of non FM states to FM ordering state because  the temperature rise at low temperatures gives rise to a lower critical field promoting further FM ordering.  
Furthermore, analysis of experimental data on the magnetization step in phase separated manganites La$_{5/8-x}$Pr$_{x}$Ca$_{3/8}$MnO$_{3}$ ($x=0.40$, 0.38, and 0.36) points to the role of thermal coupling as a key parameter of the instability in a dynamical system.\cite{MAC09} 

The Eu$_{1-x}$Sr$_{x}$MnO$_{3}$ ($x\sim 0.4$ for polycrystalline samples) system is a spin-glass like insulator at low temperature  because the substitution of Eu ion with smaller ion radius for La site gives rise to both narrow band width and an increase in the quenched disorder.\cite{SUN97,NA04,TO09}  It is well known that the former and latter parameters are controlled by the average value and the variance of the RE/AE ionic radii in  RE$_{1-x}$AE$_{x}$MnO$_{3}$, respectively.\cite{TO04} 
In  a previous work on single crystalline Eu$_{1-x}$Sr$_{x}$MnO$_{3}$ ($x\sim 0.5$), the CO phase is not formed in a long-range manner due to an increased quenched disorder. 

In this paper, to examine the steplike metamagnetic transition and its associated lattice change in the phase separated manganite, we demonstrate the effect of pressure on both the isothermal magnetization and magnetostriction in (Eu$_{1-x}$Gd$_{x}$)$_{0.6}$Sr$_{0.4}$MnO$_{3}$ ($x=0$ and 0.1). 
In \S 2, the experimental outline is described. In \S  3, the magnetization and magnetostriction data under pressure are shown with X-ray diffraction measurement on the parent sample under magnetic field. 
Section 4 is devoted to the discussion followed by the final section of summary. 

\section{EXPERIMENT}
Polycrystalline samples of  (Eu$_{1-x}$Gd$_{x}$)$_{0.6}$Sr$_{0.4}$MnO$_{3}$ ($x=0,0.1$) were prepared with a solid-state reaction method. The   stoichiometric mixtures of  Eu$_{2}$O$_{3}$, Gd$_{2}$O$_{3}$, SrCO$_{3}$, and Mn$_{3}$O$_{4}$ high purity powders were calcined in air at 1000 $^{\circ}$C for 24 h and 1250  
$^{\circ}$C for 48 h. The products were then  ground and pressed into cylindrical pellets.  The pellets were finally sintered  at $1350$ $^{\circ}$C for 36 h. X-ray diffraction data revealed that all samples are almost single phase with orthorhombic structures ($Pbnm$). 
The lattice parameters of the parent sample ($x=0$) are $a=5.4424$ \AA, $b=5.4329$ \AA, and $c=7.664$ \AA, which is in fair agreement with a previous work.\cite{SUN97} 
X-ray powder diffraction (XRD) measurements under high fields up to 4 T at low temperatures were performed with Cu$K\alpha $ radiation at the High Field Laboratory for Superconducting Materials, Institute for Materials Research, Tohoku University.\cite{WATA98} 
In particular, the XRD data at $58^{\circ }\leq 2\theta \leq 72^{\circ } $ were taken with a step size of $0.002^{\circ }$.

The electrical resistivity was measured with a four-probe method.  The magnetization measurement was carried out using commercial superconducting quantum interference device (SQUID) magnetometers both at Iwate Univ. and National Institute for Materials Science.
Hydrostatic pressures in magnetization and magnetostriction measurements were applied by using a clamp-type CuBe cell up to 1.2 GPa. Fluorinert was used as a pressure transmitting medium. The magnitude of pressure was calibrated by the pressure dependence of the critical temperature of lead. Magnetostriction was measured by using a superconducting magnet at the High Field Laboratory for Superconducting Materials, Institute for Materials Research, Tohoku University.

\section{RESULTS}
\subsection{Magnetization}
\begin{figure}[ht]
\includegraphics[width=10cm]{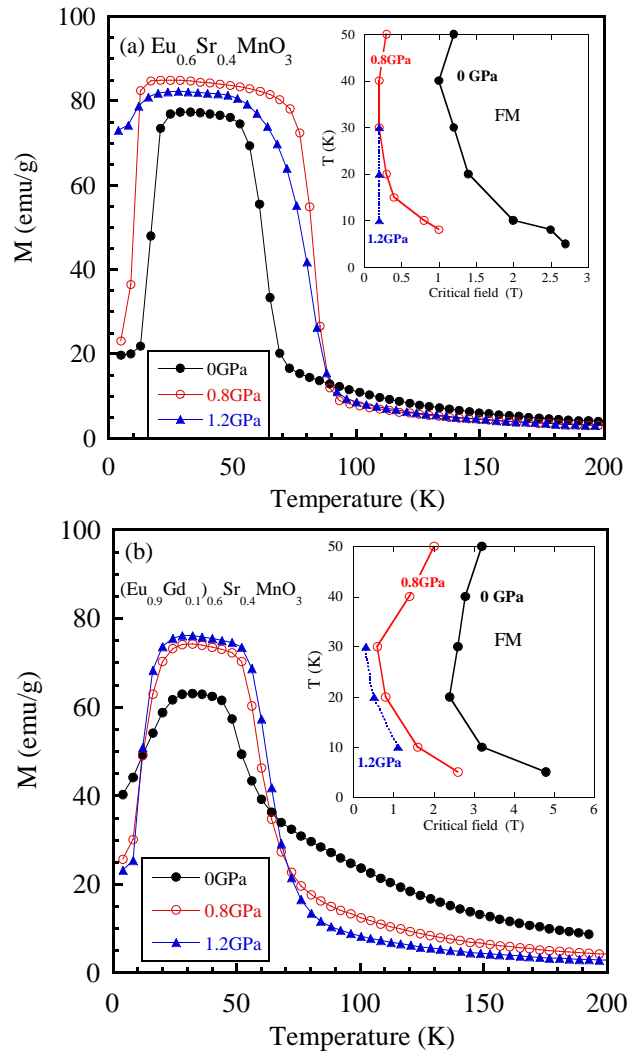}
\caption{(color online) (a) Temperature variation of the zero field cooled (ZFC) magnetization of the Eu$_{0.6}$Sr$_{0.4}$MnO$_{3}$ bulk sample under 0 GPa, 0.8 GPa, and 1.2 GPa. The $TH$ phase diagram of the parent  sample is established from isothermal magnetization data after zero field cooling at selected temperatures as shown in the inset.
The applied field is adjusted not to attain a critical field corresponding to the occurrence of a steplike behavior except for the 1.2 GPa data.
$H_{a}=$1.5 T at 0 GPa, and 1 T both at 0.8 and 1.2 GPa. 
(b) ZFC magnetization curves of the (Eu$_{0.9}$Gd$_{0.1}$)$_{0.6}$Sr$_{0.4}$MnO$_{3}$ bulk sample under 0 GPa, 0.8 GPa, and 1.2 GPa. $H_{a}=$ 3T at 0 GPa, 1.5 T at 0.8 GPa, and 1 T at 1.2 GPa. 
For comparison, the critical field versus temperature boundary is plotted as a function of the applied pressure in the inset.
 }\label{MT}
\end{figure}%

\begin{figure*}[ht]
\begin{tabular}{c}
 \begin{minipage}{0.5\hsize}
 \includegraphics[width=10cm]{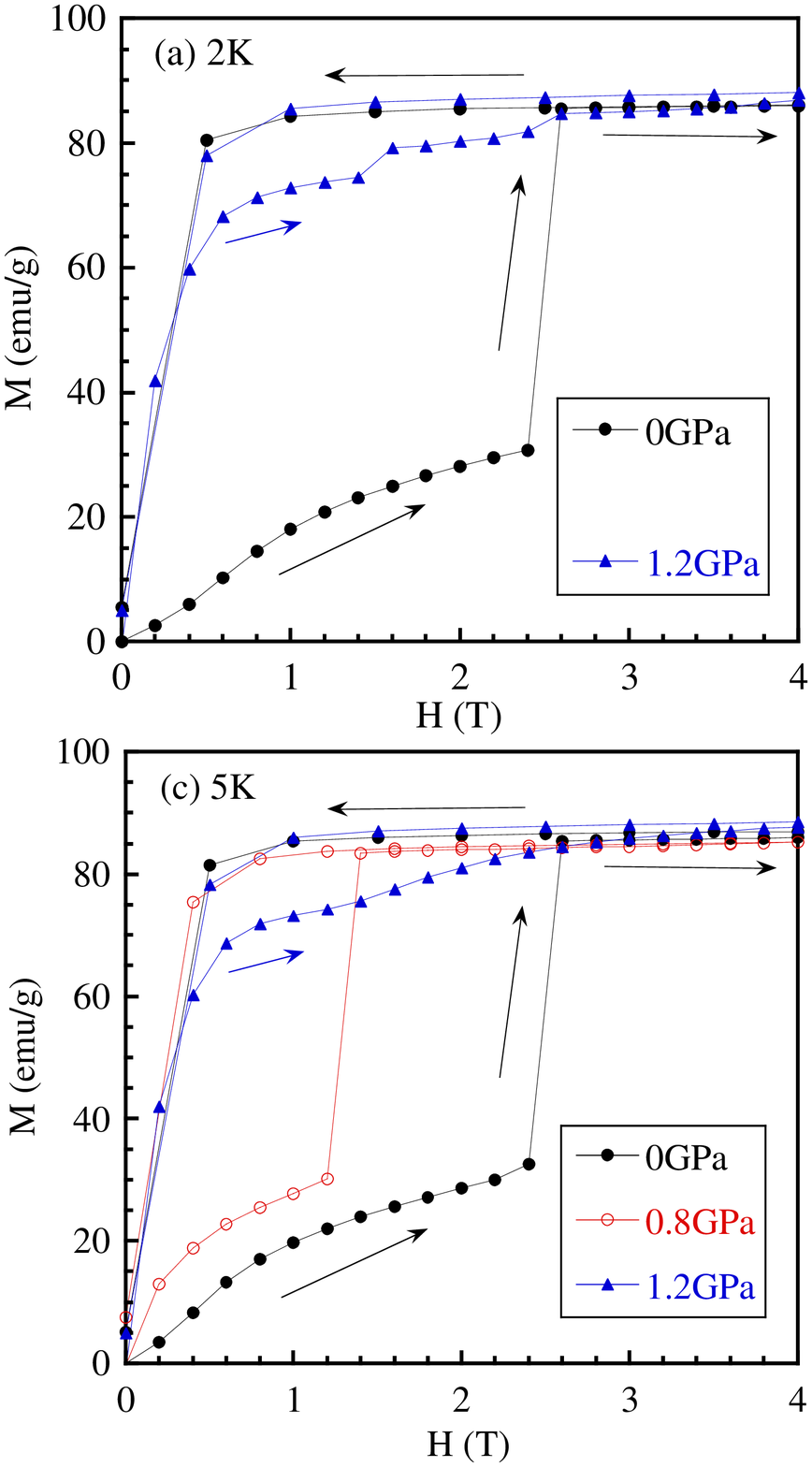}
 \end{minipage}
 \begin{minipage}{0.5\hsize}
  \includegraphics[width=10cm]{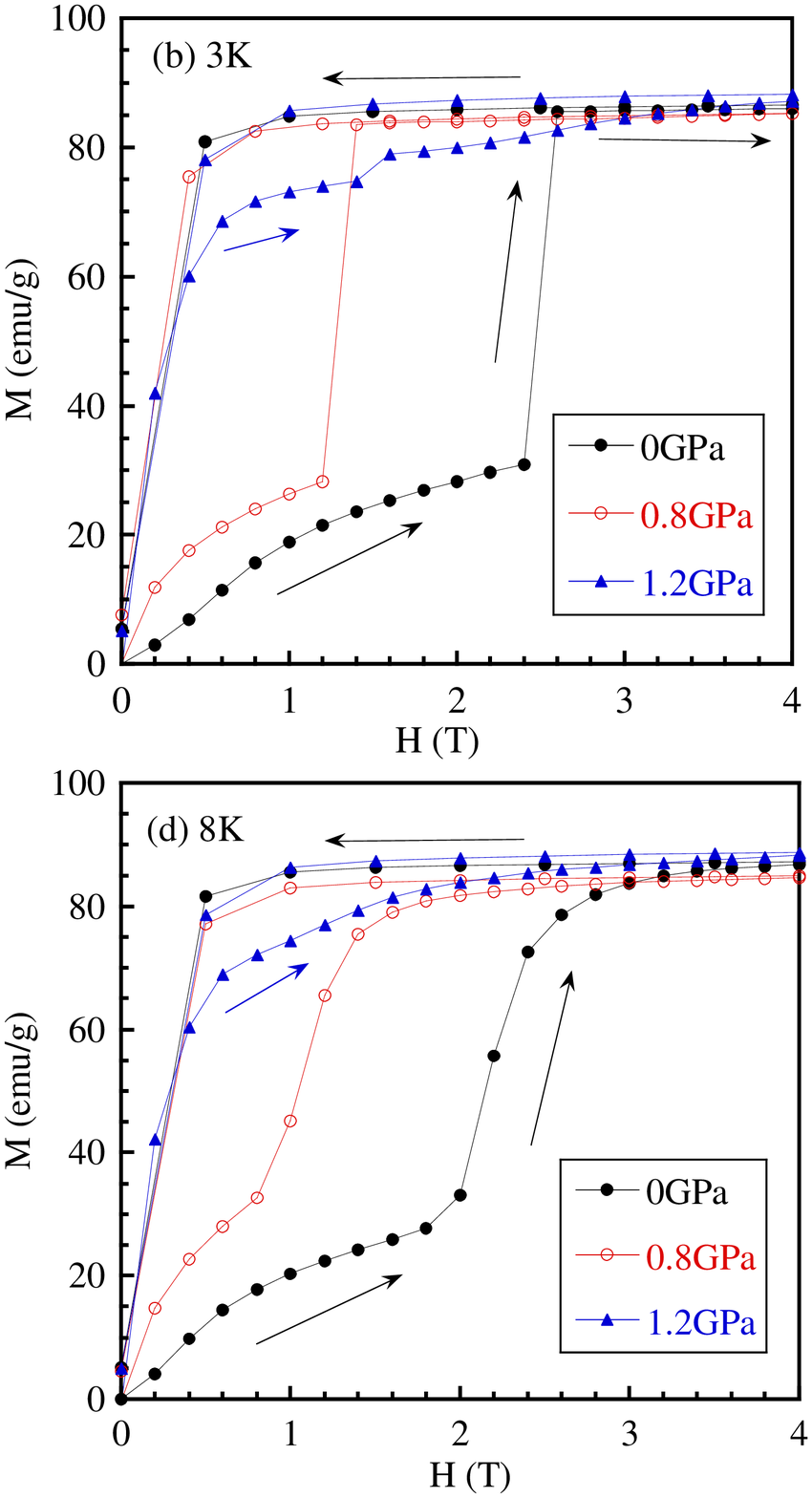}
 \end{minipage}
 
 \end{tabular}

\caption{(color online) Isothermal magnetization of the Eu$_{0.6}$Sr$_{0.4}$MnO$_{3}$ bulk sample measured under ambient pressure  and  hydrostatic pressures  at (a)2 K, (b)3 K, (c)5 K, and (d)8 K , after zero field cooling down to selected temperatures. For each measurement, a magnetic field is applied up to a maximum field at the sweep rate of 0.2 T/step. 
}\label{MH}
\end{figure*}%

\begin{figure*}[ht]
\begin{tabular}{c}
 \begin{minipage}{0.5\hsize}
 \includegraphics[width=10cm]{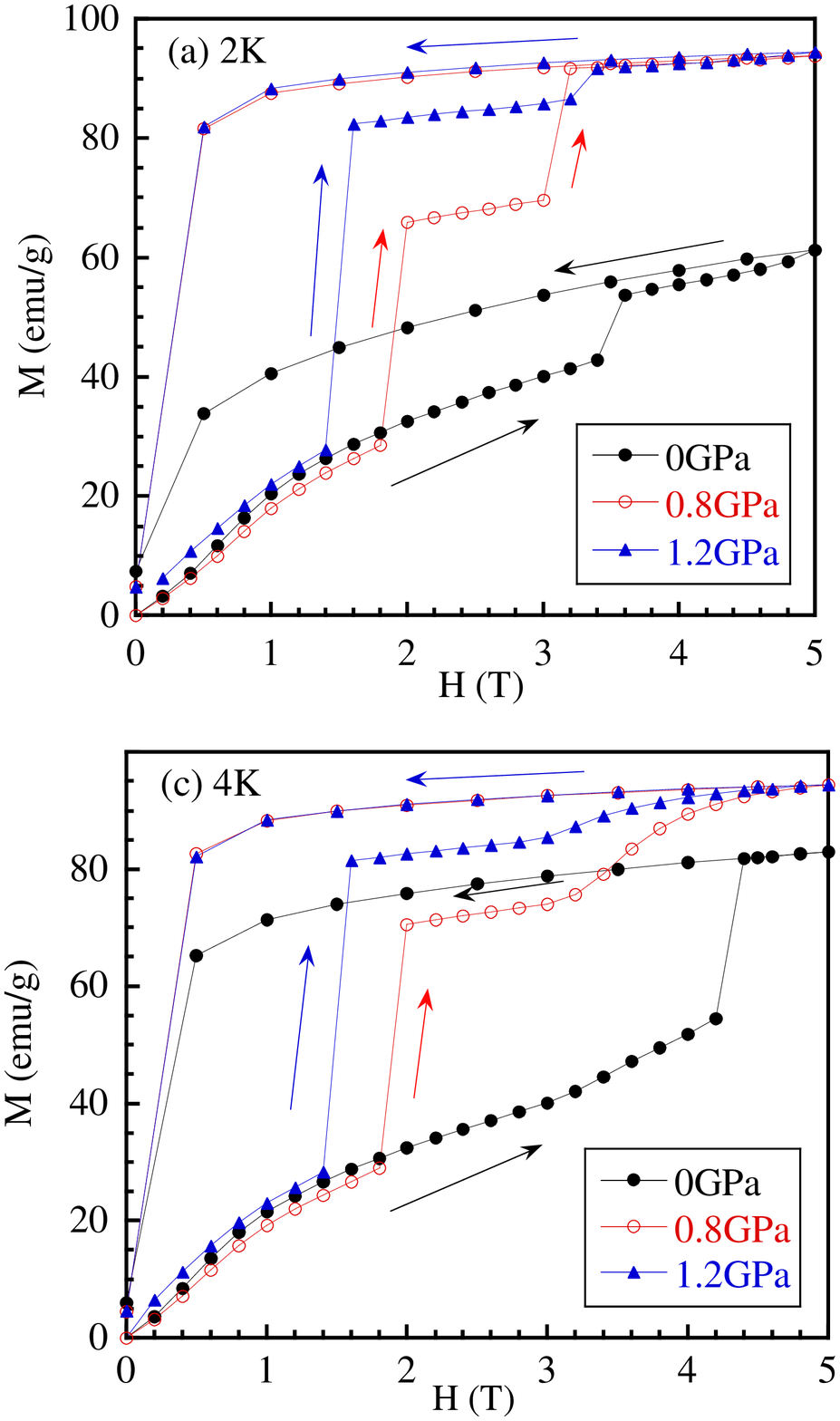}
 \end{minipage}
 \begin{minipage}{0.5\hsize}
  \includegraphics[width=10cm]{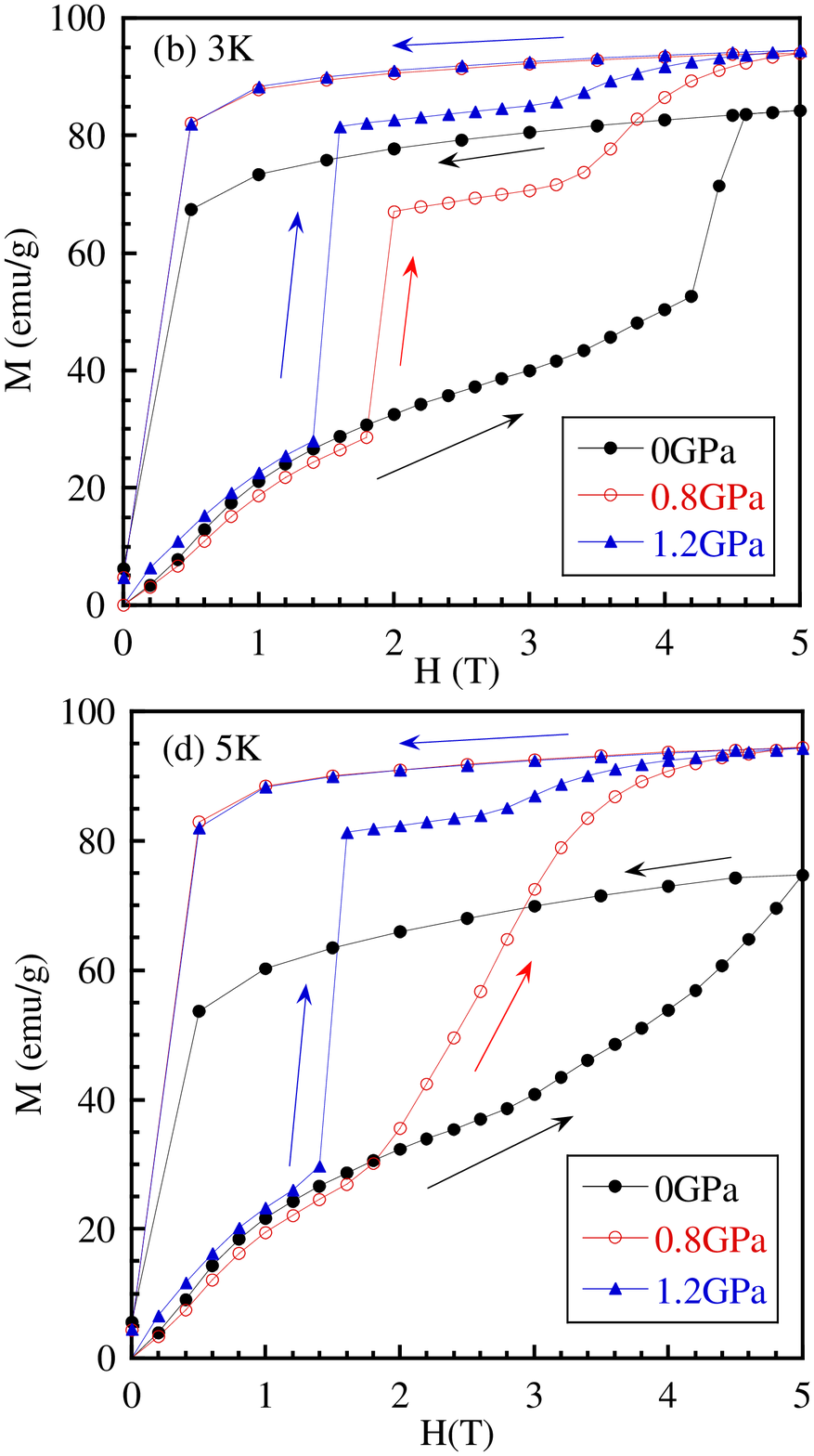}
 \end{minipage}
 
 \end{tabular}
 
\caption{(color online) Isothermal magnetization of the Gd substituted Eu$_{0.6}$Sr$_{0.4}$MnO$_{3}$ bulk sample  taken under 0 GPa, 0.8 GPa, and 1.2 GPa, after the sample was zero field cooled down to the desired temperatures of (a)2 K, (b)3 K, (c)4 K, and (d)5 K. The sweep rate is set to be 0.2 T/step up to 5 T.
}\label{MHG}
\end{figure*}%

First of all,  let us show in Fig.\ref{MT} the temperature variation of the zero field cooled (ZFC) magnetization of  (Eu,Gd)$_{0.6}$Sr$_{0.4}$MnO$_{3}$ under ambient and applied pressures.
The critical field $H_{C}$  at the corresponding temperatures is determined from the inflection points in isothermal magnetization curve upon increasing the applied field up to 5 T as shown in the inset of Fig. \ref{MT}. 
The negative curvature in $dH_{c}/dT$ at low temperatures is an essential condition for the appearance of the steplike magnetic transition as discussed in ref.\cite{GHI04,KA09}. 
As displayed in Fig.\ref{MT}(a), the ZFC magnetization curve shows a PM to FM transition near 60 K followed by a plateau, upon decreasing $T$, it then decreases rapidly until $T\sim $20 K, and finally remains a lower value down to 5 K.  We expect that the rapid suppression of magnetization observed at low temperatures is caused by a magnetic frustration between double-exchange and super-exchange interactions.   

The application of hydrostatic pressure on the ZFC magnetization causes an enlargement of the FM region, limiting a lower magnetization region below 10 K.  A volume fraction of the FM phase under pressure  is estimated to be  almost 97  $\%$ at the intermediate temperatures. For further applied pressure of 1.2 GPa, a lower magnetization  region disappears and instead of it the FM phase still remains down to 4 K, indicating no steplike transition of the Eu based sample. Moreover, we attempt to measure the ZFC magnetization curve of the Gd substituted sample, to examine the influence of chemical pressure on the magnetic properties. 
For comparison, the critical field versus temperature boundary is plotted as a function of the applied pressure in the inset of Fig.\ref{MT}(b). We note that  the $HT$ phase boundary line of the Gd10$\%$ sample under 0.8 GPa is similar to that of the parent sample at ambient pressure.

Next, to further examine the influence of lattice on the magnetic property of the bulk Eu$_{0.6}$Sr$_{0.4}$MnO$_{3}$, we attempt to measure the low-temperature magnetization as a function of field up to 5 T under ambient and applied  pressures.
At low temperatures, the steplike magnetic transition appears just around 2.6 T upon increasing the applied field as shown in Fig.\ref{MH}. The application of external pressure on the magnetization depresses the switching field for the steplike transition from  2.6 T at 0 GPa down to $\sim$1.2 T at 0.8 GPa. Furthermore, the ultrasharp phase transition to the FM state  disappears under an applied pressure of 1.2 GPa accompanied by a rapid suppression of magnetic hysteresis.  This finding is well consistent with the low temperature $MT$ curve showing no suppressed magnetic state.

To elucidate a close relationship between the lattice and spin coupling in the Gd substituted sample, we try to measure the effect of pressure on the magnetization curve of (Eu$_{0.9}$Gd$_{0.1}$)$_{0.6}$Sr$_{0.4}$MnO$_{3}$. 
Upon lowering temperature down to 4 K, a steplike magnetic transition appears around 4.2 T under ambient pressure as shown in Fig.\ref{MHG}(c). At 5 K, an abrupt transition observed at the parent sample is smeared to a continuous variation of the $MH$ curve for the  Gd10$\%$ sample (Fig.\ref{MHG}(d)). The critical field corresponding to the metamagnetic transition is determined to be 4.8 T from the inflection of the $MH$ curve. The magnitude of the magnetization is not saturate even at the maximum field of 5 T. 
The application of external pressure on the Gd10$\%$ sample lowers the switching field for occurrence of steplike transition as depicted in Fig. \ref{MHG}. At a lowest temperature of 2 K, 
a multi steplike transition occurs under the applied pressures of 0.8 and 1.2 GPa.  A second step observed around 3 T under 0.8 GPa is suppressed under the application of higher pressure of 1.2 GPa  associated with an enhanced first steplike transition,  but a tiny steplike variation remains. 
This trend is very similar to the $MH$ curve of polycrystalline Pr$_{0.5}$Ca$_{0.5}$Mn$_{0.97}$Ga$_{0.03}$O$_{3}$ with two abrupt magnetization jumps followed by plateaus.\cite{HAR03}

\begin{figure}[h]\includegraphics[width=10cm]{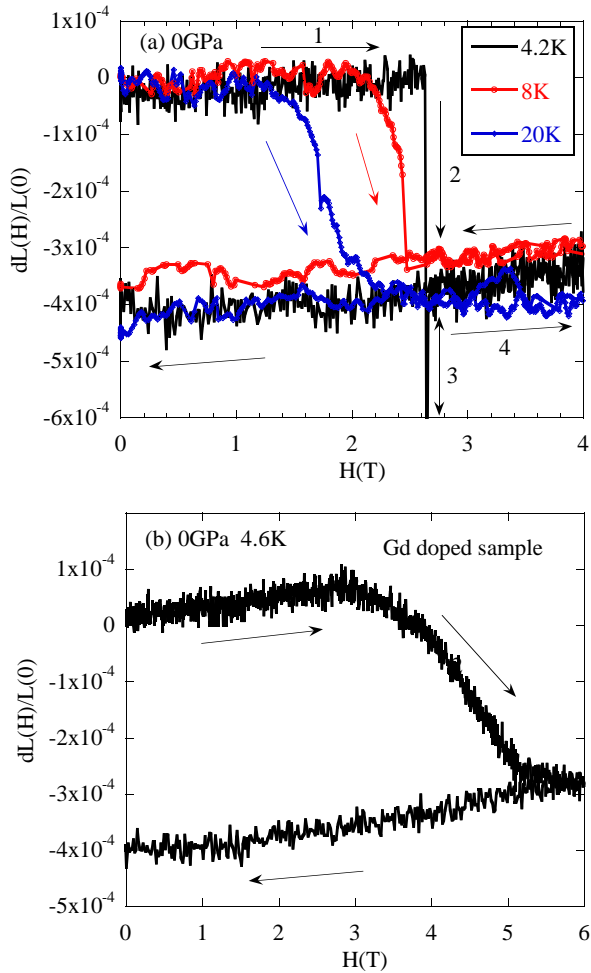}
\caption{(color online) (a)Isothermal magnetostriction $dL(H)/L(0)$ of Eu$_{0.6}$Sr$_{0.4}$MnO$_{3}$ under ambient pressure at selected temperatures of 4K, 8K, and 20K after zero field cooling.
The arrows labeled 1 and 4 denote $dL(H)/L(0)$ at 4.2 K, before and after the steplike transition.
The arrows 2 and 3 represent the net and apparent variations in $dL(H)/L(0)$ at 4.2 K.
(b) Isothermal magnetostriction of  (Eu$_{0.9}$Gd$_{0.1}$)$_{0.6}$Sr$_{0.4}$MnO$_{3}$ under ambient pressure at 4.6 K.
}\label{LH}
\end{figure}%
\begin{figure}[ht]
\includegraphics[width=10cm]{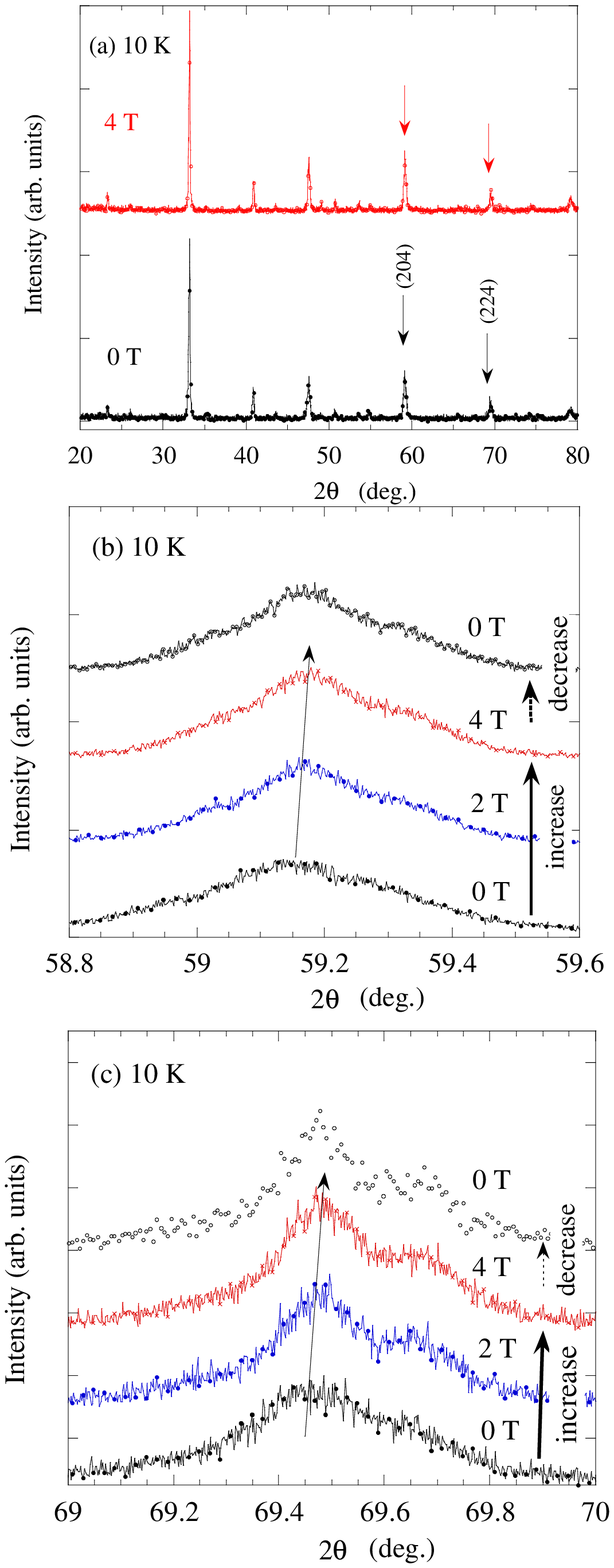}
\caption{(color online) (a)X-ray powder diffraction profiles of a powder Eu$_{0.6}$Sr$_{0.4}$MnO$_{3}$ in magnetic fields at 10 K after zero field cooling. The arrows point to the two peaks corresponding to the Miller indices (204) and (224), respectively.  (b)Field dependence of the  X-ray reflection peak located near  $2\theta =60 ^{\circ }$  recorded at 10 K  upon increasing $H$ up to 4 T and then decreasing it down to zero. (c)Field dependence of the peak around $2\theta =70 ^{\circ }$.  
}\label{Xray}
\end{figure}%


\begin{figure}[ht]\includegraphics[width=10cm]{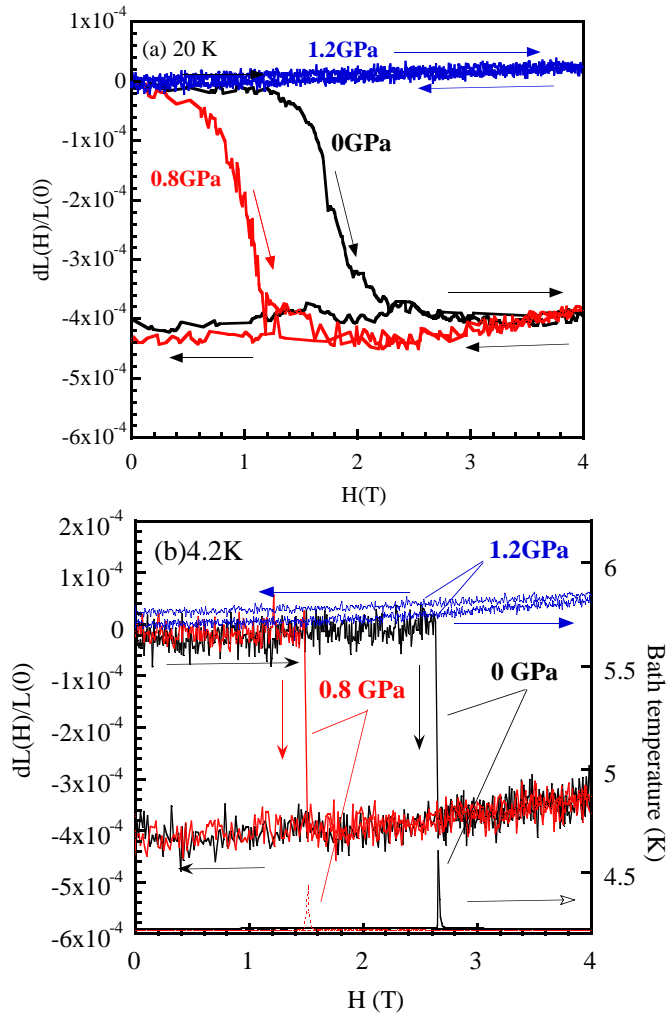}
\caption{(color online) Isothermal magnetostriction $dL(H)/L(0)$ of the parent sample as a function of the applied pressure at selected temperatures of both (a) 20 K and (b) 4.2 K. 
In (b), the  bath temperature is monitored as a function of the external magnetic field under ambient and hydrostatic pressures.  We note that the apparent change in $dL(H)/L(0)$ is removed. 
}\label{LHP}
\end{figure}%

\subsection{Magnetostriction}
 We attempt to measure the corresponding magnetostriction $dL(H)/L(0)$ of the parent Eu$_{0.6}$Sr$_{0.4}$MnO$_{3}$under ambient pressure at selected temperatures of 4 K, 8 K, and 20 K after zero field cooling(Fig.\ref{LH}). For comparison, the $dL(H)/L(0)$ data of the Gd doped sample are presented. 
We note that the magnitude of volume magnetostriction $dL(H)/L(0)=\sim-4\times 10^{-4}$ at high fields is almost independent on the direction of applied field (not shown here).  The value of $dL(H)/L(0)$ associated with the field induced FM metal transition  is in good agreement with a previous data of Eu$_{0.55}$Sr$_{0.45}$MnO$_{3}$.\cite{AB04} The steplike magnetostriction is observed upon raising the magnetic field, accompanied by the steplike metamagnetic transition.  A spiky  variation observed in $dL(H)/L$  beyond $ \sim-4\times 10^{-4}$ is not real one, which originates from by an apparent change in the resistance of strain gauge itself due to a rapid release of latent heat associated with the field induced steplike transition. \cite{MA07} 
We roughly estimate the apparent decrease in the resistance of the strain gauge to be $\sim 0.08\%$ at 4.2 K, giving a sudden temperature rise of the sample above about $\delta T\sim  $20 K, which is not very far from the value reported by Ghivelder et al.\cite{GHI04}
For the Gd substituted sample, the magnetostriction $dL(H)/L$  also corresponds to the $MH$ curve at 5 K. 

To check the effect of magnetic field on lattice parameters,  we carry out x-ray diffraction measurements of the parent sample under magnetic fields up to 4 T at 10 K as shown in Fig. \ref{Xray}. 
Our X-ray powder diffraction data show no structural phase transition associated with the field induced magnetic transition of the parent sample.
Instead, the field dependence of the peak profiles in Fig. \ref{Xray} reveals a stable decrease of the lattice parameters upon increasing applied fields. Applying the Bragg equation to the present data and differentiating it with respect to diffraction angle, we obtain a relationship that the average lattice variation $\Delta a_{0}/a_{0}$ is proportional to the peak shift $ -\Delta \theta $ in $ 2\theta $ scan. Here, we assume that $a\approx  b\approx c/\sqrt{2}= a_{0}$.  
In an applied field of $H$=4 T,  $\Delta a_{0}(H)/a_{0}(0)$ reach $-2.3\times 10^{-4} $ and $-3.4\times 10^{-4} $, for the peak shifts towards higher angle near $2\theta =60 ^{\circ }$ and $70 ^{\circ }$ , respectively, as presented in Fig.\ref{Xray}(b) and (c).
These findings qualitatively agree with the magnetostriction  behavior of $dL(H)/L$ of Fig. \ref{LH}.    

Figure \ref{LHP} shows the effect of external pressure on the magnetostriction of the parent sample which is well consistent with that on the isothermal magnetization.  In particular, the value of $dL(H)/L$ under an application of 1.2 GPa  remains almost zero upon increasing the applied field up to 7 T and exhibits no magnetic hysteresis, indicating a spontaneous transition of the system to the ferromagnetic state before the application of a magnetic field. Our data are in fairly agreement with previous results that a negative magnetovolume effect disappears above 1.15 GPa.\cite{ETO01} 
 
\section{Discussion}
First of all, the temperature dependence of the magnetization of both parent and Gd10$\%$ samples  under ambient and applied pressures, except for the high pressure data of the parent sample, is strongly suppressed at low temperatures, where is characterized by the phase separation state. On the other hand, the further application of pressure up to 1.2 GPa on the parent one causes a collapse of the low temperature PS state associated with a suppressed magnetization, resulting in the disappearance of the steplike transition. 
 The application of the external pressure on both samples lowers the critical field for the FM transition but does not change  the negative curvature in the $HT$ phase diagram, $dH_{c}/dT<0$.  Accordingly, it is concluded from our findings that the low-temperature phase separated state and its related negative temperature dependence of $H_{c}(T)$ are essential condition for the appearance of the steplike magnetic transition.
 
Here, we comment some similarities and differences in the PS state between  the La$_{5/8-x}$Pr$_{x}$Ca$_{3/8}$MnO$_{3}$ ($x=0.4$) (LPCMO) and present Eu$_{0.6}$Sr$_{0.4}$MnO$_{3}$ (ESMO). 
In previous works\cite{GHI05,SAC06}, it is believed that after zero-field cooling procedure the manganite system showing steplike magnetic transition stays in a low-temperature phase separated state with a small, and almost time independent fraction of FM regions dispersed within a CO matrix.
In other words, the low temperature PS state of the LPCMO system is taken as a blocked one preventing a growth of FM phase over the CO one in a finite time scale. On the other hand,  upon increasing temperature, the intermediate-temperature PS state above 20 K accompanies the time evolution of the FM phase, resulting in an increased magnetization  (unblocked PS state). 
We notice that a similarity in the temperature-magnetic field phase diagram including the $MT$ curve  between polycrystalline LPCMO and ESMO. 
In fact, the magnitude of ZFC magnetization under 1T in Fig.\ref{MT} is strongly suppressed at low temperatures below 20 K, which is responsible for the  blocked PS state separated from the FM one by high energy barriers. 
Effect of external pressure on the present system lowers the energy barriers separating coexisting phases and makes the blocked state unstable, resulting in a lower critical field needed for the occurrence of the steplike transition.
In the ZFC curve measured at 0.8 GPa, the low temperature PS state is limited at lower temperatures below 10 K in comparison to the magnetization data without pressure. However, it should be noted that the PS state of the present samples is composed of the FM clusters embedded within paramagnetic matrix since the temperature dependence of the magnetization exhibits such no clear CO and/or AFM transition above Curie temperature as LPCMO. 
In LPCMO system, upon decreasing temperature the CO phase is stable in a long range manner below $T_{co}\sim 230$ K, then the FM phase starts to order near $T_{c}\sim 80$ K and finally the PS state is formed below 20 K through the FM and CO states.  In the PS state of the LPCMO system, the CO matrix coexists with FM clusters, while for the EuSrMnO system the FM and CO clusters are dispersed within the PM matrix.\cite{TO09}  
This type of PS state with the PM matrix is also common to the (La$_{0.4}$Pr$_{0.6}$)$_{1.2}$Sr$_{1.8}$Mn$_{2}$O$_{7}$ bilayered manganite, where a field-induced steplike transition appears at low temperatures.\cite{MA07} 

Next, we estimate a tolerance factor, $t$, using the averaged ionic radius of A site cations ($r_{A }$ =$(1-x)r_{RE}+xr_{AE}$), which is related to the effective one-electron bandwidth of the manganite sample.\cite{HWA95}  In addition to it, we evaluate the variance of the ionic radii, $\sigma^2=\Sigma (x_{i}r^2_{i}-r^2_{A})$, where  $x_{i}$ and  $r_{i}$ are the fractional occupancies and the effective ion radii of cations of RE$^{3+}$ and AE$^{2+}$, respectively.\cite{ROD96} For the parent sample, we obtain $t=0.9183$ and $\sigma ^2=8.66\times 10^{-3} $ \AA$^2$. For the Gd substituted sample,  $t=0.9180$ and $\sigma ^2=8.79\times 10^{-3} $ \AA$^2$.  Theses finding indicate that light Gd substitution for Eu gives rise to a slight decrease in $t$ and a stable increase in  $\sigma^2$, indicating a reduction of bandwidth and an increase of the magnitude of disorder.  In a previous work\cite{KU97} on electronic phase diagram against the external pressure for the single crystalline (Nd$_{1-y}$Sm$_{y}$)$_{0.5}$Sr$_{0.5}$MnO$_{3}$ ($y=0.875$), the conversion coefficient from the external pressure to the tolerance factor is calculated to $4.6\times 10^{-4 }$/GPa. Using this pressure coefficient,  the applied pressure of 0.8 and 1.2 GPa corresponds to an increase of $t$ = $3.7\times 10^{-4 } $ and $5.5\times 10^{-4 }$, respectively.  The difference in $t$ between the parent and Gd$10\%$ samples is decreased due to the stronger applied pressure on the latter sample. 
It is true that the magnetization behavior  of the Gd$10\%$ sample under 1.2 GPa is quite similar to that of the parent one at 0.8 GPa  except for the higher field region. 
We expect that the different profile at higher fields such as the multi steplike behavior (Fig.\ref{MHG}(a)) arises from the difference in the quenched disorder between the parent and Gd doped samples. 

The application of external pressure on the magnetostriction of the sample studied lowers the critical field, but the steplike behavior in $dL(H)/L$ at 4.2 K is not changed under a high pressure of 0.8 GPa, as well as the effect of pressure on the magnetization, as shown in Fig.\ref{LH}.
Both the magnetization and magnetostriction data point to the steplike volume shrinkage associated with the ferromagnetic magnetic transition, which does not agree with the previous measurement on polycrystalline Sm$_{1-x}$Sr$_{x}$MnO$_{3}$.\cite{FIS04}  We suppose that a strain gauge attached with the sample will be  not responsible for a rapid variation although the reason is not made clear. 
It is interesting to note that a steplike lattice transformation of single crystalline (La$_{0.4}$Pr$_{0.6}$)$_{1.2}$Sr$_{1.8}$Mn$_{2}$O$_{7}$ (LPSMO)  bilayered manganite is established by the application of magnetic field. 
Accordingly, it is reasonable that the steplike magnetic transition accompanies  the ultrasharp lattice deformation at the same critical field. 

Finally, we give some comments on the effect of pressure on the magnetostriction in a bilayered manganite
because it seems to be contrary to that found in present work.\cite{YA08} 
Following the neutron diffraction study of the LPMSO bilayered manganite crystal under hydrostatic pressure,\cite{GUK05}
the applied pressure lowers the critical field of the field-induced FM states but broads the PM to FM transition region.  The external pressure strengthens the DE interaction in the bilayers, while it weakens the super exchange  interaction between the adjacent bilayers.  This finding is close to a different response of crystal structure to the applied pressure between bilayered and cubic manganites.
In addition, we note that in previous works on the effect of pressure on the transport, magnetization, and magnetostriction of Eu$_{0.58}$Sr$_{0.42}$MnO$_{3}$ compound by several authors, novel phenomena such as steplike magnetic transition and/or its associated magnetostriction are never referred.\cite{ETO01,KAJ06,ZHA07}

\section{SUMMARY}We have investigated the influence of external pressure on the steplike magnetic transition and  its concomitant magnetostriction  in the polycrystalline (Eu,Gd)$_{0.6}$Sr$_{0.4}$MnO$_{3}$. 
A suppression of the energy barrier separating between the FM and low-temperature phase separated states under pressure leads to lowering the critical field. 
The further application of external pressure up to 1.2 GPa on the parent sample substantially weakens the low temperature phase separated state, resulting in collapses of the ultrasharp metamagnetic transition and its concomitant lattice striction.
X ray diffraction patterns of the parent sample taken under high fields at low temperatures show the stable decrease of average lattice parameters, which is consistent with the magnetostriction behavior.   
Our data indicate the importance of the low-temperature phase segregated state in the  doped manganite system exhibiting the steplike transition. 

 This work was partially supported by a Grant-in-Aid for Scientific Research from Japan Society of the Promotion
of Science. 

\end{document}